*nanomaterials*

MDPI

# Ultrathin Washcoat and Very Low Loading Monolithic Catalyst with Outstanding Activity and Stability in Dry Reforming of Methane

**Fazia Agueniou, Hilario Vidal \*, María Pilar Yeste, Juan C. Hernández-Garrido, Miguel A. Cauqui, José M. Rodríguez-Izquierdo, José J. Calvino and José M. Gatica**

Departamento de Ciencia de Materiales e Ingeniería Metalúrgica y Química Inorgánica, e IMEYMAT, Instituto Universitario de Investigación en Microscopía Electrónica y Materiales, Universidad de Cádiz, 11510 Puerto Real, Spain; fazia.agueniou@alum.uca.es (F.A.); pili.yeste@uca.es (M.P.Y.); jcarlos.hernandez@uca.es (J.C.H.-G.); miguelangel.cauqui@uca.es (M.A.C.); josemaria.izquierdo@uca.es (J.M.R.-I.); jose.calvino@uca.es (J.J.C.); josemanuel.gatica@uca.es (J.M.G.)
* Correspondence: hilario.vidal@uca.es; Tel.: +34-956012744

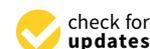

**Abstract:** A Ni/CeO$_2$/ZrO$_2$ catalyst with improved redox properties has been washcoated onto a honeycomb cordierite monolith in the form of a nonconventional alumina-catalyst layer, just a few nanometers thick. In spite of the very low active phase loading, the monolith depicts outstanding performance in dry reforming of methane, both in terms of activity, with values reaching the thermodynamic limit already at 750 °C, even under extreme Weight Hourly Space Velocities (WHSV 115–346 L·g$^{-1}$·h$^{-1}$), as well as in terms of stability during prolonged Time on Stream (TOS 24–48 h).

**Keywords:** nickel; ceria-zirconia; dry reforming of methane; honeycomb cordierite; FIB-(S)TEM

## 1. Introduction

Supported nickel catalysts stand out currently as the most promising candidates for the valorization of carbon dioxide into syngas by catalytic Dry Reforming of Methane (DRM). This is a process which attracts increasing interest due to its potential to contribute to the abatement of two of the gases with highest greenhouse potential, CO$_2$ and CH$_4$ [1]. Moreover, DRM is also a suitable process for the utilization of the biogas produced by anaerobic digestion of waste biomass, which consists mainly of CH$_4$ and CO$_2$ [2]. Since Ni-based catalysts are prone to deactivate by deposition of carbon residues, the use of oxide supports with high oxygen handling capacity, such as ceria-zirconia [3] and ceria-zirconia-yttria [4], has been proposed to oxidize the carbon deposits under operating conditions.

In most works, these catalytic formulations keep being applied in fixed bed reactors, [5–7] despite the well-documented advantages (low pressure drop, extended heat and mass transfer properties, and large geometrical surface area), and fruitful application of structured catalysts in the field of Environmental Catalysis, in particular as washcoated monoliths [8]. Moreover, the honeycomb monolithic design allows optimizing the amount of catalytic active phase. In the particular case of both Ni and 4f-elements which are currently classified as critical raw materials [9], this is really attractive as a way to guarantee their efficient usage, in per mass terms.

Although some studies reported on the use of structured catalysts on the methane steam and oxy-steam reforming [10], as well as biogas autothermal reforming [11], references that employed such design in DRM are much scarcer and none of them compared the same proposed catalyst in a powder form and supported onto a honeycomb monolith [12]. Therefore, this gap will be faced in this study, which additionally explores the beneficial effects of using a very low loading of a nickel-ceria based formulation in DRM.

## 2. Materials and Methods

Following the above guidelines, a $Ni/CeO_2/ZrO_2$ catalyst exhibiting improved redox properties was deposited by washcoating onto a cordierite honeycomb type monolith. Details about the preparation of both the starting powder catalyst and the monolithic devices are described in the Supplementary Materials section. In brief, different monoliths pieces (230 cpsi) were washcoated with a suspension of the catalyst (19.1 wt%) into a polyvinyl alcohol (1.7 wt%), colloidal alumina (4.2 wt%), and water. The residual slurry, containing the catalyst, was dried and submitted to calcination at 450 °C (1 h) to obtain the reference powdered sample for the catalytic study. This powdered sample had 3.9 wt% of Ni content and a 0.18 Ce/Zr molar ratio as determined by means of ICP-AES analysis. The catalyst loading onto the cordierite surface (0.36 mg·cm$^{-2}$) was estimated from the weight gain of the coated monoliths. The coating adherence was evaluated from the weight loss after immersion of the monoliths in petroleum ether under ultrasounds (30 min).

Scanning Electron Microscopy (SEM) results were acquired on a Thermofisher Nova NanoSEM 450 microscope (Eindhoven, The Netherlands) whereas Scanning Transmission Electron Microscopy (TEM/STEM) data were recorded in a Thermofisher Titan$^3$ Themis 60–300 double aberration-corrected microscope (Eindhoven, The Netherlands) working at 200 kV. Both microscopes allowed complementary Energy Dispersive Spectroscopy (EDS) compositional analysis. Electron-transparent lamellas obtained by Focused Ion Beam (FIB) for STEM were prepared using a Zeiss Auriga Dual Beam microscope (Oberkochen, Germany).

Catalytic activity tests in the DRM reaction for a 1:1 mixture of pure $CH_4$ and $CO_2$ were carried out in the temperature range 750–900 °C, under Weight Hourly Spatial Velocities (WHSV) from 115 up to 346 L·g$^{-1}$·h$^{-1}$, and for Time on Stream (TOS) values between 24 and 48 h. The latter were selected to operate under steady-state conditions considering the endothermic character of the reaction, so avoiding differences in temperature transients between powdered catalytic beds and monolithic reactors. More details can be found in the Supplementary Materials section.

## 3. Results and Discussion

Figure 1 shows the results of a study by Scanning Electron Microscopy (SEM) of the distribution of the catalyst over the monolith walls. After observing different areas of several coated monoliths, two important features are worth being highlighted: (1) The catalyst grains are efficiently spread out over the available surface, in the form of a discontinuous, patchy like, layer, Figure 1A. Though their size is varied, no large accumulation of the active phase takes place. Moreover, as revealed by the X-ray Energy Dispersive (EDS) maps, Figure 1B–E, the grains correspond to the $Ni/CeO_2/ZrO_2$ powder; (2) the three elements in the active phase keep in tight contact after deposition by washcoating onto the monolith. Particularly, Ni is well dispersed over the ceria-zirconia support.

Nevertheless, conventional SEM imaging does not provide information about the structure of the washcoating layer. To this end, an electron transparent cross-section of the monolith surface was prepared by FIB techniques and characterized by STEM, using High Angle-Annular Dark-Field mode (HAADF), Figure 2A,B, following the fruitful approach previously reported [13,14].

Figure 2A shows a low-mag view of the FIB cross-section (about 10 μm) of the monolith surface. Note first, the large roughness of the monolith, in the order of a few microns. The bright contrast part of the STEM-HAADF image corresponds to the Au and Pt protecting layers deposited during the preparation of the FIB lamella [15].

Likewise, the lower intensity region of the image corresponds to the cordierite support. The washcoating layer is so thin that it cannot be appreciated at such low magnification. It is nevertheless clearly observed in the high magnification zoom of the area marked in Figure 2A, as shown in Figure 2B. In this image a porous structure is evidenced over the compact area corresponding to the cordierite support. The thickness of this zone is just tens of nanometers (≈50 nm), a value close to the size of the catalytic particles themselves, which appear embedded, anchored, within its porous structure, and exposed at the topmost layers.

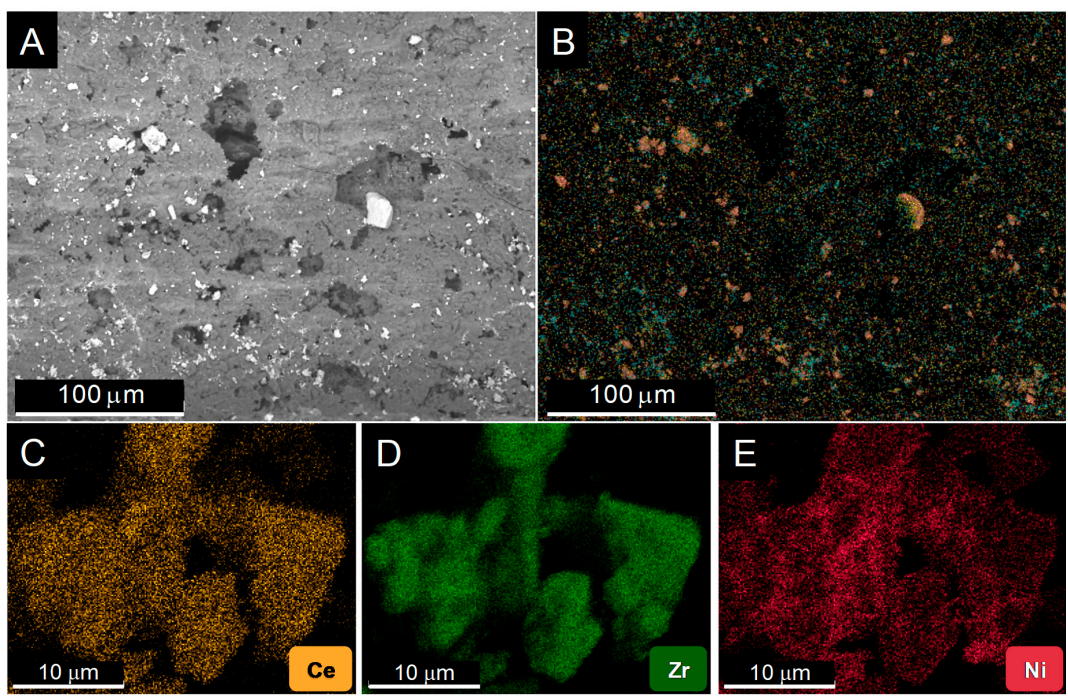

**Figure 1.** Representative secondary electron Scanning Electron Microscopy (SEM) image of the monolith inner walls (**A**) and the corresponding composed SEM-EDS map overlaying the chemical maps of the different elements in the active phase (**B**). Individual maps for Ce (**C**), Zr (**D**), and Ni (**E**) were obtained from a more reduced area of the sample.

Such nanometric thick alumina-catalyst washcoat is orders of magnitude thinner than those commonly found in conventional preparations of monoliths [16], and clearly provides a route to avoid burial of the active phase particles at positions a few microns beneath the surface, far below the contact region with the reactants flow. Therefore, the design here reached represents a clear step forward in terms of efficiency in the usage of the active phase, an important issue in the case of critical raw materials.

Moreover, the nanoporous nature of the washcoat provides means, not only to fix the active phase particles to the monolith surface, but also to guarantee access of the reactants to the surface of the active phase particles.

To confirm the structure of the washcoat layer described above, a nanoanalytical study by STEM-EDS was also performed using the FIB sample in an Aberration Corrected FEI Titan[3] Themis 60–300 microscope, Figure 2C. Note how the Si and Al signals are present, as expected, in the bottom zone corresponding to the cordierite. In this area, a Ti signal due to an inclusion in the ceramic material is also detected. On its hand, the Al signal is also present in the upper layer but in the form of a porous structure. This is the area previously identified as corresponding to the washcoat.

It is also evident that the washcoat adapts to the hill and valley morphology (roughness) of the cordierite surface, this surely providing a better anchoring of the coating layer onto the monolith walls. This feature was confirmed by the resistance to mechanical stress during the adherence test from which a value of 92% was estimated.

The Al signal in the washcoat layer decreases only in the area occupied by a particle of the active phase (top right). The chemical maps indicate that Ce distributes mostly by wrapping the zirconia crystallite. Ni, in the form of particles with sizes ranging from 10 to 30 nm, is in contact with the ceria-zirconia support, in good agreement with the SEM-EDS observations.

Figure S1 in the Supplementary Materials section shows the spectra corresponding to the whole area of the catalyst particle. The peaks corresponding to the different elements in the active phase (Ce, Zr, and Ni) are clearly identified.

　　　　Finally, the Pt and Au signals appearing at the top are due to the protecting layers deposited to prevent sample damage during the preparation of the FIB lamella.

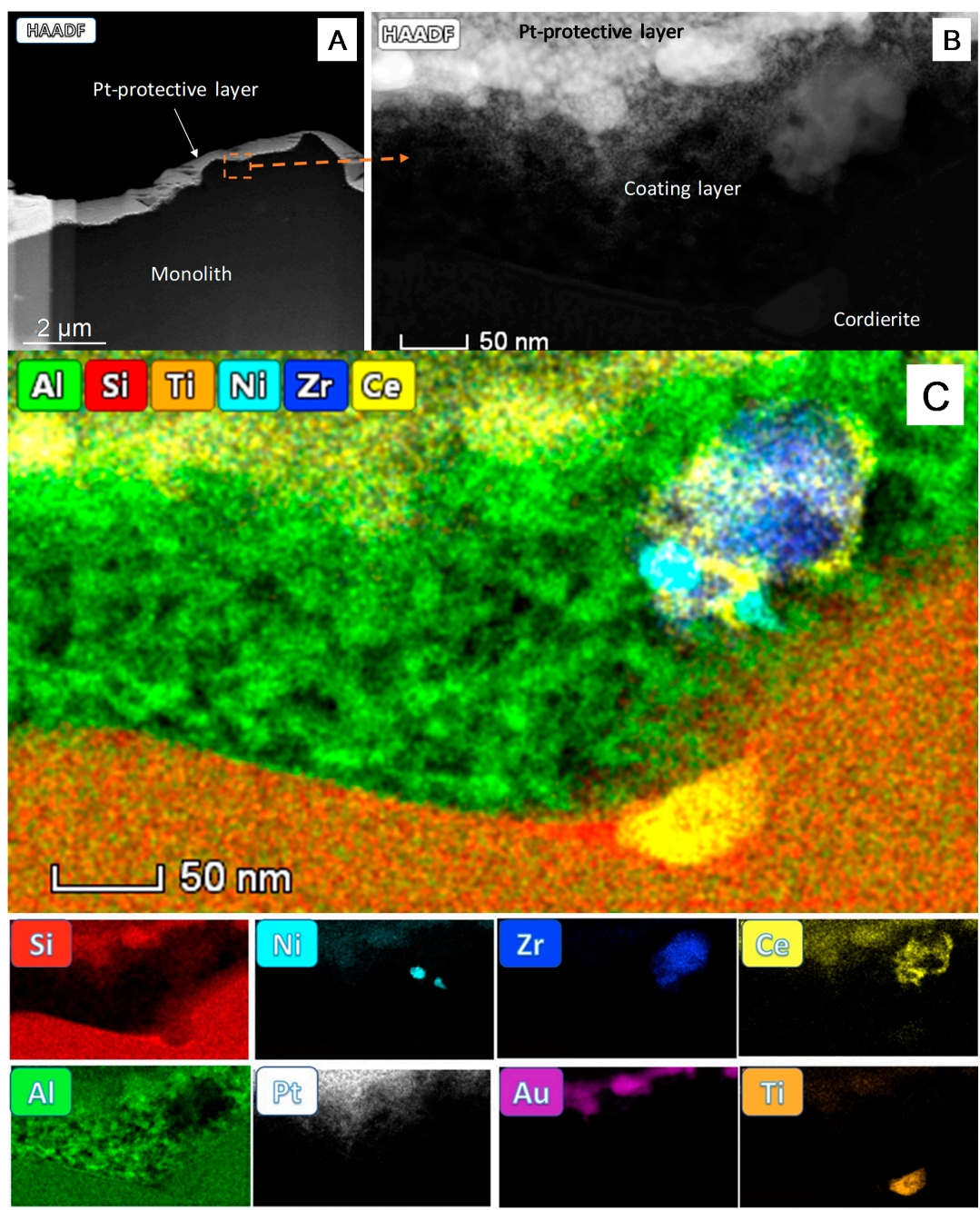

**Figure 2.** Scanning-Transmission Electron Microscopy in High Angle-Annular Dark-Field mode (STEM-HAADF) image of the Focused Ion Beam (FIB) cross-section of the washcoating layer (**A**). A zoom of the marked area is shown in (**B**) and the STEM-EDS analysis of this area (**C**). The map overlaying all the elemental signals in the catalytic device is shown at the upper part of (**C**) while the individual maps corresponding to the above elements besides Pt and Au used during the sample preparation are gathered below.

　　　　The results of catalytic tests are gathered in Figure 3, which compares the performance of the monolith with that of the powder catalyst. Note first the large effect of depositing the catalytic formulation onto the ultrathin washcoat monolith (Figure 3A). For the latter, $CO_2$ conversion values higher than 90% and reaching the thermodynamic limits (Table S1 in Supplementary Data section) are observed above 750 °C. In the powder, conversions at all temperatures are far below those of the monolith, especially at the lowest temperature (750 °C). This comparison between monolith and powder is also favorable to the first if the activity is expressed as $CO_2$ consumed per unit of time and Ni mass, with values of 13.6 and 7.9 mmol/s·g Ni, respectively, at 750 °C and WHSV of 115 L/g·h.

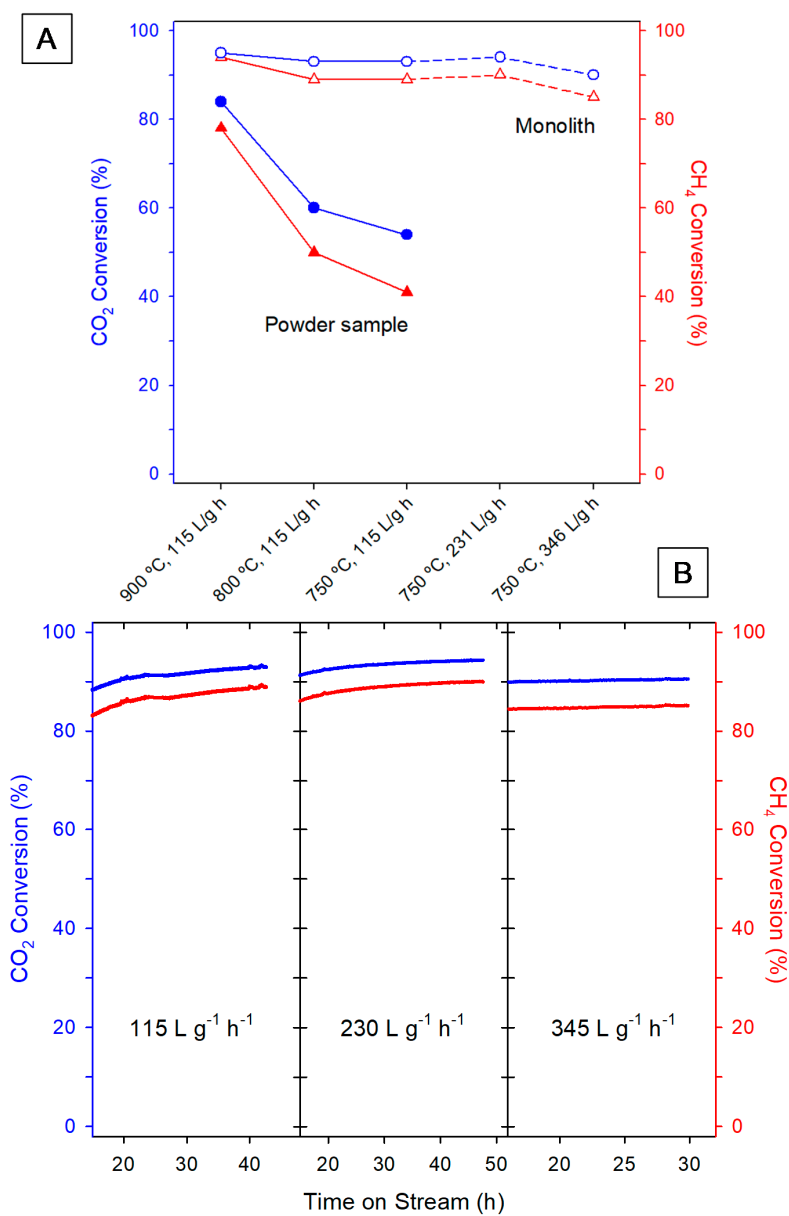

**Figure 3.** Catalytic performance results: (**A**) $CO_2$ (blue circles) and $CH_4$ (red triangles) conversions after 24 h Time on Stream (TOS), at different temperatures and Weight Hourly Space Velocities (WHSV) values for different Ni/CeO$_2$/ZrO$_2$ samples in the form of monolith and powder; (**B**) evolution with TOS of the $CO_2$ (blue plots) and $CH_4$ (red plots) conversions at 750 °C and increasing WHSV values for different monolithic samples.

It is also worth noting that increasing WHSV from the lower value, 115 L·h$^{-1}$·g$^{-1}$, up to 345 L·h$^{-1}$·g$^{-1}$ does not significantly influence the products conversion in the monolithic device, which behaves quite steadily. It is noticeable that such WHSV values are quite high when compared with those usually employed in the literature [17], not only for this reaction but in general in papers dealing with the use of catalyst monoliths. Therefore, the monoliths prepared in this work perform exceptionally well. In addition, it is clear that the monolith configuration exerts a quite positive influence on the performance of the powder catalyst, providing advantages with respect to the fix bed reactor.

At this point, it should be highlighted that this remarkable performance at all temperatures and WHSV values is obtained with a monolith containing a very small specific loading of the washcoat (0.36 mg·cm$^{-2}$), a value at least one order of magnitude lower than in conventional monoliths [18]. A positive effect of decreasing the catalyst loading on DRM has been recently reported in a study using a cordierite-based monolith, in which a Co-Ru-Zr active phase was added by wet impregnation of the metallic precursors [19]. Nevertheless, the WHSV values investigated in this case fall far below those used here. Moreover, a reference Co-Ru-Zr catalyst is missing in this contribution, which precludes any conclusion about the effect of dispersing the same catalytic formulation onto the monolith surface.

As stated in the first paragraph of the introduction, a quite important issue in DRM refers to the stability of the catalyst and, in turn, of the catalytic reactor under working conditions. At this respect, the plots in Figure 3B clearly indicate that both $CO_2$ and $CH_4$ conversion remain stable in the case of the monoliths, both under low and high flow conditions, even after prolonged reaction periods. The $H_2$/CO ratio is also an important output in DRM reaction, values close to 1 being desired to allow an easier adaptation to many downstream processes including ammonia and methanol synthesis [17]. On this regard, we obtained $H_2$/CO ratio data ranging from 0.85 to 0.89 for all the studied honeycomb samples.

## 4. Conclusions

A honeycomb cordierite loading very small amounts of a Ni/CeO$_2$/ZrO$_2$ catalyst was prepared with quite unconventional structural and catalytic performance in dry reforming of methane. Very small aggregates of the active phase, in which the different components of the formulation are kept in tight contact, remain well spread within a highly porous ultrathin alumina layer, which wets the rough surface of the monolith walls. This, jointly with the intrinsic activity of the starting powder catalyst, explains the behavior of the structured device, even operating at low temperatures (750 °C) and high reactants flows, and the absence of deactivating carbonaceous species. The high activity that reaches the thermodynamics limits, and the stability of our monoliths in the studied reaction have been obtained with an unusual catalyst low loading suggesting that the increase of this synthetic parameter is not necessary.

**Supplementary Materials:** The following are available online at http://www.mdpi.com/2079-4991/10/3/445/s1. Synthesis of the powdered and monolithic catalysts, experimental details of the catalytic test, and thermodynamic estimate. Figure S1: Quantitative compositional analysis results from a selected area of the catalyst shown in Figure 2 including Ni-Ce-Zr elements and their corresponding EDS spectra. Elemental contents should be taken with caution considering that they correspond to a tiny portion of the washcoat. Table S1: $CO_2$ and $CH_4$ equilibrium conversion data for the DRM reaction conducted at 1 atm and with $CH_4$:$CO_2$ = 1:1 based on thermodynamic analysis.

**Author Contributions:** Conceptualization and methodology, H.V., M.A.C., J.J.C., and J.M.G.; investigation, F.A., M.P.Y., and J.C.H.-G.; resources, J.M.R.-I.; writing—original draft preparation, J.C.G.; writing—review and editing, H.V.; supervision, H.V. and J.M.G.; funding acquisition, M.A.C. and J.J.C. All authors have read and agreed to the published version of the manuscript.

**Funding:** This research was funded by the Ministry of Economy and Competitiveness of Spain/FEDER Program of the EU (Project MAT2017-87579-R) and Junta de Andalucía (Groups FQM-110 and FQM-334).

**Acknowledgments:** Electron microscopy data were recorded at the facilities of SC-ICYT (UCA).

**Conflicts of Interest:** The authors declare no conflict of interest.